\documentclass[a4paper]{article}

\usepackage{amsmath}
\usepackage{graphicx}   
\usepackage[american]{babel}
\usepackage[utf8]{inputenc}
\usepackage[T1]{fontenc}
\usepackage{afterpage}
\usepackage{color}		
\usepackage{epsfig}
\usepackage[font={footnotesize}]{caption}
\usepackage{lineno}

\def\hyp{{\hbox{-}}}

\begin{document}

\title{Joint optimization of replication potential and information storage set the letter size of primordial genetic alphabet}

\author{Hemachander Subramanian\\
\normalsize{Department of Physics}\\
\normalsize{National Institute of Technology, Durgapur, W.B., India} \\
\normalsize{hemachander.subramanian@phy.nitdgp.ac.in} } 
\date{\today}
\maketitle

\begin{abstract}
The simplest possible informational heteropolymer requires only a two-letter alphabet to be able to store information. The evolutionary choice of four monomers in the informational biomolecules RNA/DNA or their progenitors is intriguing, given the inherent difficulties in the simultaneous and localized prebiotic synthesis of all four monomers of progenitors of DNA from common precursors on early Earth. Excluding the scenario where a two-letter alphabet genome eventually expanded to include two more letters to code for more amino acids on teleological grounds, we show here that a heteropolymer sequence in the RNA-world-like scenario would have had to be composed of at least four letters in order to predictably fold into a specific secondary structure, and hence must have outcompeted the two-letter alphabet genomes. Using a model that we previously used to demonstrate the evolutionary advantages of unidirectional replication and anti-parallel strand orientation of duplex DNA, we show here that the competing constraints of maximum replicative potential and predictable secondary structure formation can be simultaneously satisfied only by palindromic heteropolymer sequences composed of a minimum of four letters, within the premise of the presence of sequence-dependent asymmetric cooperativity in these RNA/DNA progenitors.  
\end{abstract}

\section*{Statement of Significance}
Is there any evolutionary significance for the letter size of the genetic alphabet? When replication rate become sequence-dependent, conflicts arise between the choice of sequences that maximize replication rate and sequences that maximize information storage. This is resolved by increasing the number of letters to four.

\section*{Background}

The information in the biological heteropolymers DNA and RNA is encoded in four letters. DNA and RNA being products of evolution \cite{primitivegeneticpolymers,grandfathersaxe,simplernucleotidesjoyce,leslie2004prebiotic}, it is reasonable to investigate the evolutionary significance of this choice of four-letter alphabet. What was the evolutionary advantage of prebiotic informational heteropolymers composed of four letters, over heteropolymers composed of lesser number of letters? \cite{Crick1968, szathmary4letters}. Does the evolutionary force that selected four-letter heteropolymers still persist in extant genomes? The prevailing reasoning is that at least four letters are needed to code for the twenty amino acids, since, with the three-nucleotide genetic code, one can have $4^3=64$ possible amino acid letters, enough to code for twenty amino acids with degeneracies (three letters would not accomodate heteromolecular base-pairing between nucleotides of DNA/RNA). The presumed sequence of events leading to a four-letter RNA possibly involved beginning with a binary-letter RNA (or an RNA progenitor), and later acquiring two more letters (monomers) that enabled expansion of the genetic code to include more amino acids. However, this explanation does not hold up to a closer scrutiny, since, the acquisition of monomers corresponding to four letters by RNA must have come \textit{before} the eventual expansion of the genetic code by evolution, which would have rendered those RNA molecules at a severe evolutionary disadvantage in the period between the letter expansion and the genetic code expansion. The evolutionary disadvantage of those RNA molecules would stem from the need to maintain a sufficient feed of all four monomers for the RNA self-replication, which would have required the presence of more complex geochemical processes, compared to the requirements for self-replication of a binary-letter RNA, without any concomitant accrual of evolutionary benefits to the former \cite{MillerLevy,  joyce1989rna, pyrimidinesynthesis, rnaworldorigins1, prebioticchemistry}. This brings us back to the initial question about the evolutionary significance of four-letter heteropolymers, within an ``RNA-world''-like scenario, where the information stored in the heteropolymer sequences was not about protein construction, but about the secondary and tertiary folded structures of the heteropolymers that were capable of catalysing the processes involved in their own replication. Our arguments below satisfy the constraint of immediate accrual of benefit to the replicator upon expansion of the letter space, thereby avoiding teleological explanations.

Within our premise described below, both the replicative potential of a self-replicating heteropolymer strand and the specificity of its secondary structure depend on its sequence. We show that, in binary-letter heteropolymer strands, the sequences that maximize the replicative potential of heteropolymer strands lack secondary structure specificity, i.e., cannot fold into a specific, predictable secondary structure, due to the availability of many possible secondary structure configurations. Conversely, binary-letter sequences that promote specificity/predictability of secondary structures cannot have maximal replicative potential. These two conflicting demands are met simultaneously by the introduction of another pair of monomers, resulting in four-letter heteropolymer strands, that can both replicate efficiently and are able to form predictable secondary structures capable of catalyzing reactions associated with their self-replication. The evolutionary advantage stemming from the superior replicative and catalytic capabilities of four-letter heteropolymer possibly offset the disadvantages described in the previous paragraph, leading to its preference over the binary-letter heteropolymer.


\subsection*{\textbf{The Premise}: Asymmetric Cooperativity in DNA}
In our earlier investigations \cite{firstpaperjtb,scirep} exploring the evolutionary advantages of the unidirectional $5'$ to $3'$ replica strand construction of DNA single strand and the anti-parallel strand orientation of duplex DNA, our assumption of the presence of asymmetric cooperativity in DNA helped us demonstrate that both the aforementioned properties are outcomes of evolutionary maximization of replicative potential of progenitors of DNA. We also supplied substantial literature-based experimental support for the presence of asymmetric cooperativity in DNA in our earlier articles \cite{firstpaperjtb,scirep}. Although the exposition below has already been covered in greater detail in our earlier article \cite{scirep}, we include it here to keep this article self-contained,  and delay the introduction of ideas original to this article until the ``Results'' section.

\subsubsection*{Sequence-Independent Asymmetric Cooperativity}
\begin{figure}
\begin{center}
\includegraphics[width=0.9\textwidth]{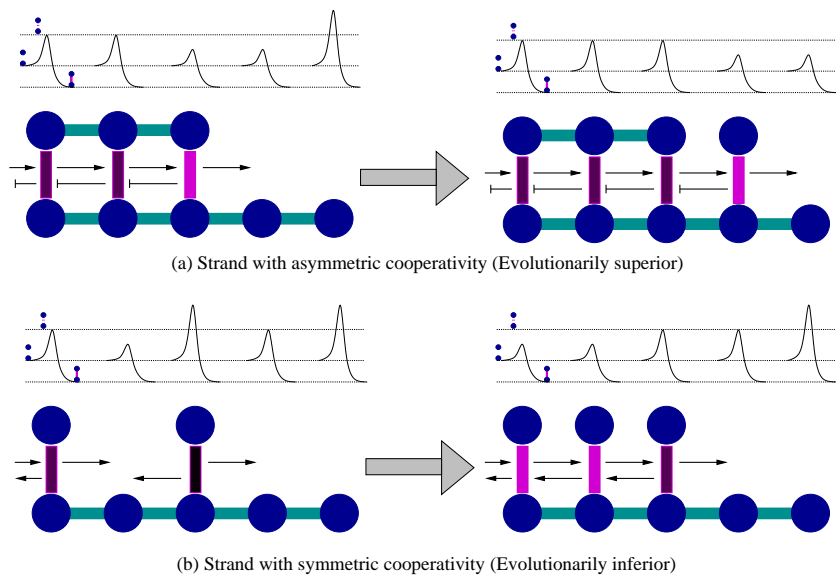}
\caption{Illustration of replica strand construction process of a circular autocatalytic polymer in the presence of (a) asymmetric and (b) symmetric nearest-neighbor hydrogen bond cooperativities (figure and caption reprinted from \cite{firstpaperjtb} with permission from Elsevier). The circles represent monomers and the thick vertical lines connecting a pair represent the inter-strand hydrogen bonds. Horizontal lines connecting the monomers represent covalent bonds between them. The color of the hydrogen bonds represent the height of the kinetic barrier separating bonded and unbonded configurations, higher the barrier, darker the color. Hydrogen bonding energy diagram is shown above for both cases of cooperativity. The bottom line of the three lines in the energy diagram corresponds to the energy of the bonded configuration, and the middle line corresponds to that of the unbonded configuration. Only a section of the circular strand is shown here for convenience. (a) In a strand with asymmetric cooperativity, the catalytic influence on and from a hydrogen bond's left and right neighbors are unequal, simplified here to be catalysis and inhibition from left and right, respectively. Catalysis from a neighboring hydrogen bond is denoted by an arrow from the neighboring bond, and inhibition, with a bar-headed arrow. Such non-reciprocal, asymmetric catalytic influence among a pair of neighboring hydrogen bonds leads to low kinetic barrier for hydrogen bond formation/dissociation near the growth front, enabling faster monomer utilization for the replica strand construction. This also prevents the pre-existing hydrogen bonds behind the growth front from dissociation and provides them longer lifetime to facilitate covalent bond formation between the replica strand monomers. This asymmetrically cooperative hydrogen bonding behavior solves the conundrum of simultaneously requiring both low and high kinetic barriers for hydrogen bond formation/dissociation stemming from the conflicting needs of high monomer utilization and high covalent bond formation probability, but renders the strands directional. (b) In strands with symmetric cooperativity, two hydrogen bonds mutually catalyze each other's formation/dissociation, by reducing the kinetic barriers symmetrically. In this case, already formed hydrogen bonds in regions away from the growth front (second bond from left) have smaller kinetic barriers than the bonds at or near the growth front (fourth bond from left). This reduces the template's ability to attract monomers for replica strand elongation, and also its ability to keep the monomers bonded long enough to facilitate covalent bond formation, thereby reducing its replicative potential relative to templates with asymmetric cooperativity. Please note that the scales of energies in (a) and (b) are not the same.}
\label{symasym}
\end{center}
\end{figure}

We define asymmetric cooperativity as unequal and non-reciprocal \textit{kinetic} influence of an interstrand hydrogen bond on its left and right neighboring hydrogen bonds, as illustrated in fig.\ref{symasym}. More explicitly, when an interstrand hydrogen bond in DNA/RNA-like heteropolymer lowers the kinetic barrier for formation and dissociation of its \textit{left} neighboring hydrogen bond, and raises the kinetic barrier of its right neighboring bond, the strand is said to be in the \textit{left} asymmetrically cooperative mode. Strands in right asymmetrically cooperative mode can be similarly defined. The presence of asymmetric cooperativity in DNA/RNA-like heteropolymer single strands increases their replicative potential by simultaneously satisfying two competing requirements for their self-replication: low kinetic barrier for interstrand hydrogen bond formation to easily induct monomers, and high kinetic barrier to retain the monomers already hydrogen-bonded to the template, enabling intrastrand covalent bond formation and replica strand elongation\cite{firstpaperjtb}. For example, during the self-replication of a right-asymmetrically cooperative, single-strand heteropolymer, a monomer hydrogen-bonded to the template strand would decrease the kinetic barrier for the formation of another hydrogen bond to its \textit{right}, thereby drawing monomers to bond closer to itself and enable covalent bond formation and daughter strand extension. On the other hand, the newly-formed second hydrogen bond to the right would increase the kinetic barrier for the \textit{dissociation} of the already-formed first hydrogen-bonded monomer to its \textit{left}, as shown in fig. \ref{symasym}, thereby increasing the bond's lifetime and hence the probability of covalent bond formation. Thus, asymmetric cooperativity increases the probabilities of both monomer induction and retention, and improves the replicative potential of the heteropolymer. The incorporation of asymmetrically cooperative kinetic interactions betwee the monomers require their two ($5'$- and $3'$-) covalent-bonding ends to be structurally distinct \cite{jmepaper}. The above-proposed asymmetric cooperativity is of \textit{kinetic origin} and goes beyond the thermodynamic directional asymmetry, well-known from the experimentally determined nearest-neighbor free energies, enthalpies and entropies of pairs of base-pairs \cite{santalucia1998}.

\begin{figure}
\begin{center}
\includegraphics[width=0.9\textwidth]{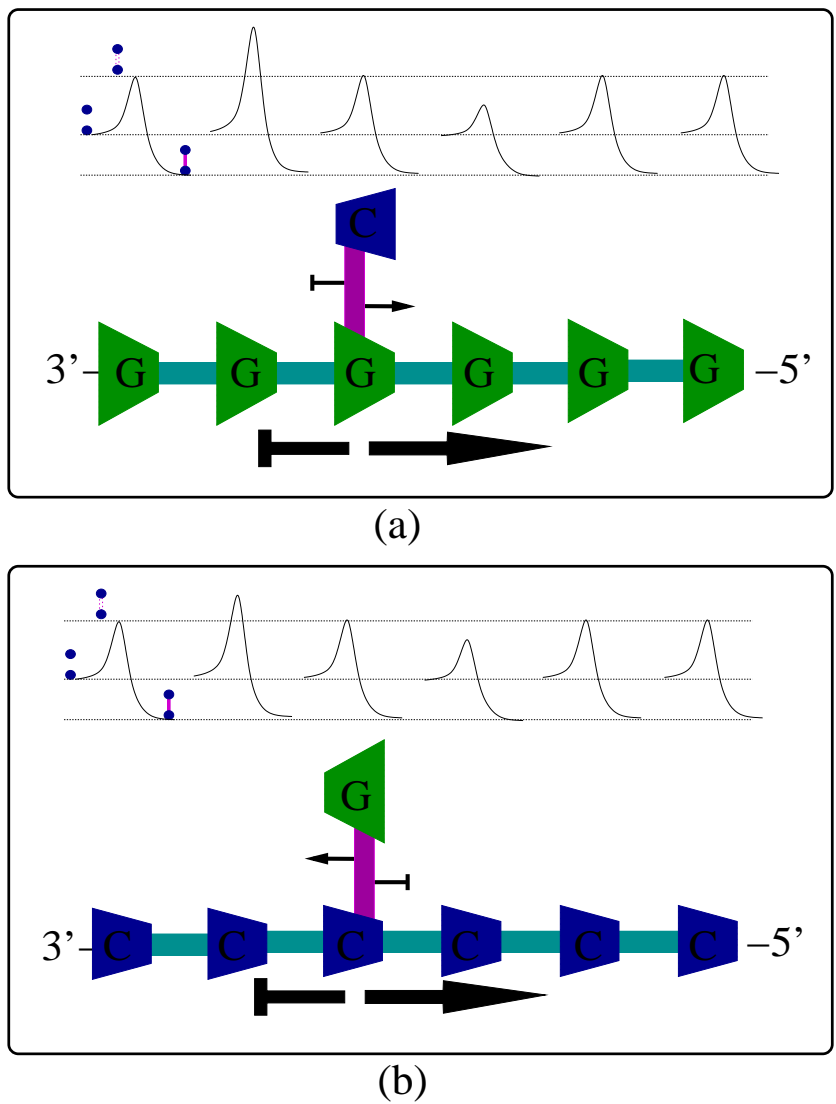}
\caption{Illustration of sequence-independent asymmetric cooperativity in DNA single strands. The asymmetric cooperativity mode of a single template strand is dictated by the $3'-5'$ directionality of the strand (figure and caption adapted from \cite{scirep} under a Creative Commons Attribution $4.0$ International License). (a) A hydrogen bond between a lone nucleotide $C$ and the template strand catalyzes the formation of another hydrogen bond to its right by reducing its kinetic barrier, while inhibiting the formation of its left neighbor by raising its barrier. The strength and the mode of sequence-independent asymmetric cooperativity dictated by the template strand is denoted by the thick black arrow below the template strand, pointing to the right. The thin arrows attached to the hydrogen bond denote the weaker sequence-dependent asymmetric cooperativity strength and mode. (b) Irrespective of the type of nucleotides occupying the template strand, the directionality of the template strand alone dictates the mode of sequence-independent asymmetric cooperativity. The thinner arrows on the hydrogen bond, denoting sequence-dependent part, though pointing in the opposite direction, does not alter the overall mode of asymmetric cooperativity due to the relative strength of sequence-independent part. This can be seen in the kinetic barrier diagrams above the bonds, where the barrier on the right of the hydrogen bond is lower in both (a) and (b), and the barrier on the left is higher. This assumption ensures that the daughter strand is always constructed from the $5'$ end of the daughter strand to its $3'$ end.}
\label{seqindependent}
\end{center}
\end{figure}

Earlier \cite{scirep}, we theoretically factorized asymmetric cooperativity into a stronger sequence-independent part operative in DNA \textit{single} strands, and a weaker sequence-dependent part operative in the DNA \textit{double} strands, in order to demonstrate the evolutionary advantage of anti-parallel DNA duplex strands and to justify the counter-intuitive evolutionary choice of the complicated, piecemeal ``lagging strand'' replication mechanism. The sequence-independent asymmetric cooperativity dictates the direction of construction of replica strands, by decreasing the kinetic barrier for new hydrogen bond formation towards one direction (the $5'$-end) of the single-stranded template strand, and raising it for the already-formed bonds behind the growth front, to stabilize them, as mentioned above. The evolutionary advantage provided by sequence-independent asymmetric cooperativity to the single strands of self-replicating progenitors of DNA rendered them (and their constituent monomers) directional \cite{firstpaperjtb}, which is reflected in the non-equivalence of $3'$ and $5'$ ends of DNA and its constituent nucleotides, within our model. The effect of the presence of sequence-independent asymmetric cooperativity on the kinetic barriers of hydrogen bonds between the template and replica strands of a self-replicating heteropolymer is illustrated in fig.\ref{seqindependent}. It has to be reiterated that the mode of the sequence-independent asymmetric cooperativity (left or right) is dictated by the $3'-5'$ directionality of the template strands themselves and is thus sequence-independent \cite{firstpaperjtb, scirep}.

The usual argument against the above asymmetric cooperativity-based explanation for the replicative directionality of daughter strand on the single strand of DNA begins by noting the $3'-5'$ asymmetry of the nucleotides themselves. Due to the presence of the activating group at the $5'$-end of the nucleotide that carries the energy for covalent bond formation, daughter strand construction can proceed only along the $5'-3'$ direction. This $3'-5'$ asymmetry materially results in the directionality of the daughter strand construction on the DNA template. However, the price the DNA pays for such unidirectional replication on the single-stranded template is quite high, since simultaneous replication of the anti-parallel DNA strands involves piecemeal replication (Okazaki fragments) of the lagging strand replicas, requiring a complex replication machinery and coordination between the replication processes of the two strands. The obvious \textit{biophysical} cause of strand directionality, mentioned above, should not preclude one from inquiring into the \textit{evolutionary} cause of the same (corresponding to the distinction between the Aristotle's material and the final causes), especially when the adoption of strand directionality appears to impose significant costs for DNA replication. Due to the supervenience of evolution over chemistry \cite{primitivegeneticpolymers,grandfathersaxe,simplernucleotidesjoyce,leslie2004prebiotic}, at evolutionary time-scales, function must dictate structure, and apparent structural constraints such as monomer asymmetry must have been overcome if it involved significant fitness costs. Identifying an evolutionary advantage for such an asymmetry might resolve the above conundrum, which was the central objective of our earlier paper \cite{firstpaperjtb}.

\subsubsection*{Sequence-Dependent Asymmetric Cooperativity}
\begin{figure}
\begin{center}
\includegraphics[width=0.9\textwidth]{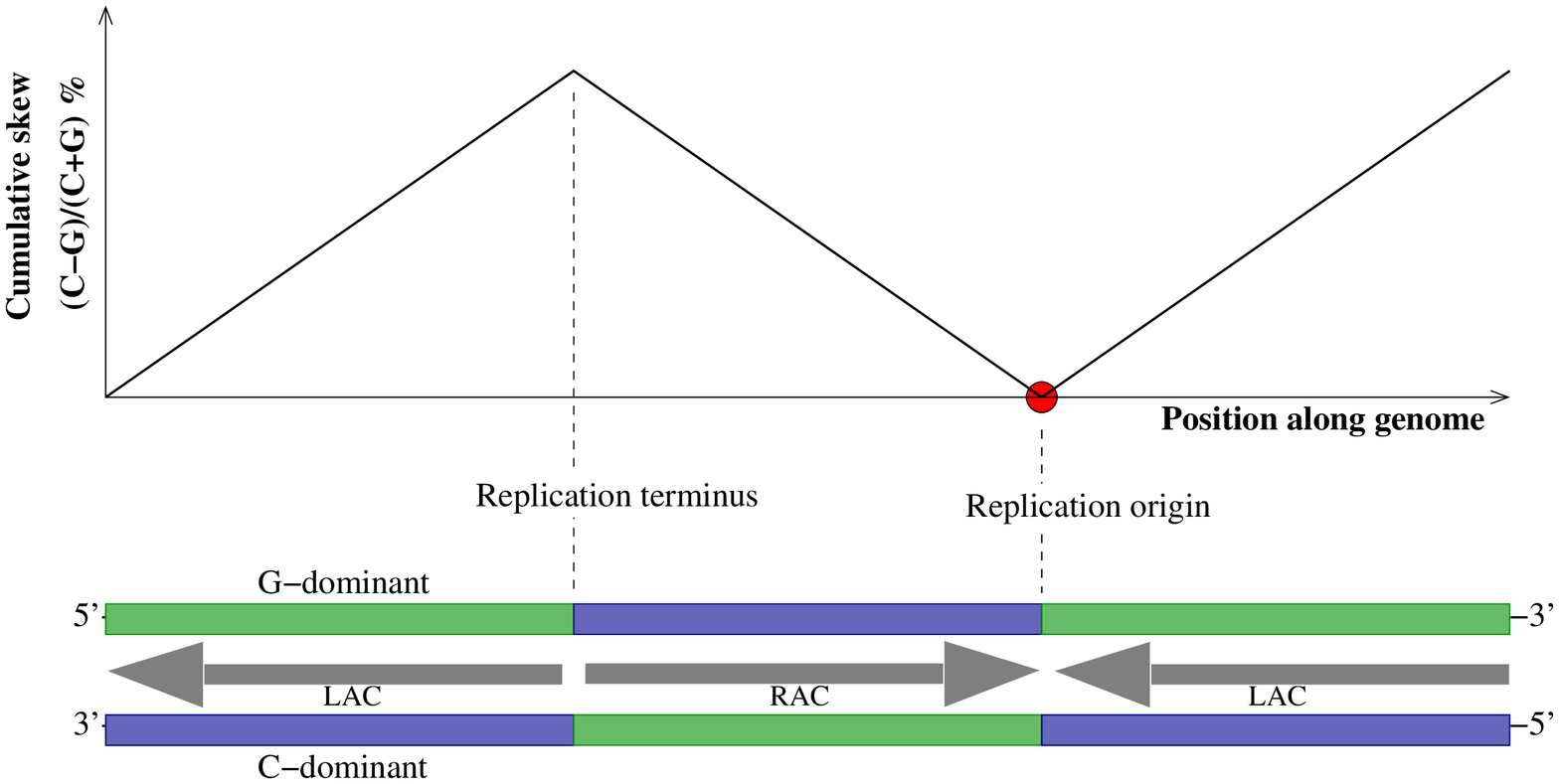}
\caption{Schematic diagram illustrating the experimental observations related to $GC$ skew in various genomes (figure and caption adapted from \cite{scirep} under a Creative Commons Attribution $4.0$ International License). The genome is composed of independently replicating subunits, usually refered to in biological literature as ``replichores'', three of which are shown here colored in blue and green. The replichore that is enriched in $C$ is denoted in blue, and in $G$, green. The sign of $GC$ skew has been observed to correlate with the direction of replication along a template strand. Leading strands, the segments where replication and unzipping machineries travel in the same direction, have been observed to be enriched in the nucleotide $G$.  Lagging strands, where these two machineries travel in opposite direction, must then be enriched in $C$. Note again that the asymmetric cooperativity mode defined here is opposite to that of the figure in \cite{scirep}. It has also been observed that only one of the boundaries between the replichores function as origin of replication, whereas the other, as replication terminus. The schematic graph above the strands illustrate the cumulative $GC$ skew, calculated in running windows of appropriate size over the entire genome. It is representative of skews observed in genomes of multiple species, and clearly shows the boundaries between replichores. The $GC$ skew has traditionally been attributed to the difference in replication mechanisms between the leading and lagging strands. All these observations are understandable within our picture of sequence-dependent asymmetric cooperativity, where $GC$ skew is treated as the \textit{cause} of unzipping directionality. The arrows between the two strands, labeled RAC and LAC, denote the right and left modes of sequence-dependent asymmetric cooperativity respectively. At the origin of replication, pointed by a red dot on the graph above, the asymmetric cooperativities reduce the barrier from both left and right, rendering the bonds at the interface weaker and thereby allowing the interface to function as the origin. At the terminus, the barrier height for the hydrogen bonds are raised from both directions and can be broken only when the neighboring bonds are broken.}
\label{gcskew}
\end{center}
\end{figure}

The weaker sequence-dependent asymmetric cooperativity dominates in DNA \textit{double} strands, since the stronger sequence-independent part, dictated by the $3'-5'$ directionality of the single strands, stands cancelled due to the anti-parallel orientation of the two strands in the duplex. The sequence-dependence of asymmetric cooperativity in duplex strands arises from the dependence of the \textit{mode} of asymmetric cooperativity on the ``orientation'' of the hydrogen-bonding base-pairs, which differentiates, for example, the base-pair $5' \hyp G \hyp 3'/3' \hyp C \hyp 5'$ from that of its $180^\circ$-rotated counterpart, $5' \hyp C \hyp 3'/3' \hyp G \hyp 5'$. Thus, the base-pair $5' \hyp G \hyp 3'/3' \hyp C \hyp 5'$ would lower the kinetic barrier for the formation/dissociation of its left hydrogen bond neighbor and raises the barrier for its right neighbor (left mode), whereas, the $180^\circ$-rotated base-pair $5' \hyp C \hyp 3'/3' \hyp G \hyp 5'$ catalyzes its right neighbor and inhibits its left neighbor (right mode). The effect of the presence of sequence-dependent asymmetric cooperativity on the kinetic barriers of hydrogen bonds in a DNA duplex is shown in fig. \ref{seqdependent}. The presence of sequence-dependent asymmetric cooperativity in DNA duplex strands renders the rate and the direction of unzipping of the duplex \textit{sequence-dependent}, thereby providing a crucial, additional degree of freedom to control the kinetics of DNA replication. We have earlier argued \cite{scirep} that this additional degree of freedom is exploited by evolution to create sequences that allow for simultaneous replication of multiple segments of DNA, called replichores in Biology literature, thereby increasing the rate of replication substantially. This is done by creating segments with alternating modes of asymmetric cooperativity, by loading the top strand in one segment with more G's, and loading it in the next segment with more C's, and so on, i.e., through \textit{asymmetric nucleotide composition} (GC and AT skews). The two types of interfaces between such segments function as origins and termini of replication, resulting in replication being carried out bidirectionally from the origins, to the left and right, enabled by the alternating permissive orientations for unzipping of the replichores. The cumulative skew diagram arising from GC skews is illustrated in fig. \ref{gcskew}. Such skews are found in nearly all genomes, both prokaryotes and eukaryotes, studied thus far, and are widely used as a bioinformatic tool to find the origins of replication \cite{rocha2004replication,replicationorientationskew,SPombeorigins,gcskeworigins1,gcskeweukaryotes,originsmammalsnatcomm,oriloc}. By selecting appropriate sequences, the number of replication origins and hence the replication rate of the DNA can be regulated at evolutionary timescales to suit the environmental constraints such as the availability of monomers \cite{divideandconquer}. Within our model, due to this evolutionary advantage provided by sequence-dependent asymmetric cooperativity, RNA and DNA's evolutionary progenitors have evolved molecular structures for monomers that supported the incorporation of sequence-dependent asymmetric cooperativity, and later bequeathed them to the former. It has to be remembered that, while the unzipping rate of the DNA \textit{duplex} depends on the sequence, daughter strand construction on the \textit{single-strand} template is relatively sequence-independent, due to our assumption of the strength of sequence-independent asymmetric cooperativity in relation to its sequence-dependent counterpart. Additionally, the sequence-dependent asymmetric cooperativity mode (left or right) for the base-pairs in \cite{scirep} was chosen primarily to demonstrate the replicative advantage of  anti-parallel strand orientation of DNA, for which the choice of direction of asymmetric cooperativity modes was unimportant. Here we define the directionality of sequence-dependent asymmetric cooperativity modes to be opposite to that of  [\cite{scirep}], to drive home the point that the two modes are entirely equivalent symmetry-broken solutions arrived at through maximization of replicative potential, and the possibility that DNA and RNA may have opposite asymmetric cooperativity modes. It is important to note that our choice of the mode of sequence-dependent asymmetric cooperativity in DNA/RNA \textit{does not affect our conclusions} about the evolutionary advantage of anti-parallel strands in \cite{scirep} or of quadruplet-letter alphabet in this article in any way, just as the choice of left or right directionality of a bag's zipper is impertinent to its functionality.

\subsubsection*{Experimental support for Asymmetric Cooperativity}
Multiple experiments in the literature, when appropriately reinterpreted, support the existence of asymmetric cooperativity in DNA. Since sequence-dependent asymmetric cooperativity reduces the kinetic barrier to the left and increases the barrier to the right (or vice versa, depending upon the mode), unzipping DNA from the left and right ends should produce distinct force signatures. Bockelmann \textit{et al} \cite{stickslipprl} observed such distinct unzipping force signatures when they unzipped a single molecule phage $\lambda$ DNA using an Atomic Force Microscope. During unzipping of double-stranded DNA using a nanopore \cite{nanoporeunzipping,nanoporeunzipping2}, White's group found that the unzipping rates are dependent on the orientation of the entry of the DNA into the nanopore. The two sequences $5'\hyp(AT)_6(GC)_6\hyp3'$  and $5'\hyp(GC)_6(AT)_6\hyp3'$ have nearly the same thermodynamic stabilities, but their unzipping kinetics have been shown to differ by orders of magnitude \cite{at6gc6sequence}. Cooperativity in DNA (un)zipping, yet another signature of neighborhood influence on hydrogen bond dissociation and formation, has been abundantly documented in the literature \cite{dnaunzippingprl2009,dnaunzippingpnas,dnacooperativityprE,dnacooperativitypnas2003,mechanicalunzippingkumar}.

Strong experimental support for the presence of sequence-dependent asymmetric cooperativity in DNA comes from the dependence of rate of extension of a template-attached primer on the neighboring nucleotides \cite{dnatemplatingKervio}. This group has measured the \textit{kinetics} of non-enzymatic extension of a primer by a single activated nucleotide, in the presence of a down-stream binding strand, and has shown that the rates of nucleotide incorporation depends on the orientation of the neighboring base-pairs. As mentioned above, due to sequence-dependent asymmetric cooperativity, for the same force of unzipping, the rate of (un)zipping from one end of the duplex would be different from the rate of (un)zipping from the other end, and depends on the sequence. This asymmetry has been observed recently in an experiment involving strand displacement reaction \cite{dnaunzippingFPT}: This group has observed that the strand displacement reaction rates are significantly different depending upon whether the strand displacement begins at the $3'$ end of the template or at its $5'$ end, and these differences are strongly dependent on the sequences. Similar differences between $3'$-end- and $5'$-end-initiated strand displacement reactions have been observed in toe-hold mediated strand displacement in RNA:DNA duplexes as well \cite{rnadnakinetics}. The presence of polar replication forks and G-enriched regions in genome, where the replication fork unzips and traverses easily along one direction in the DNA but struggles to traverse the region in the opposite direction, strongly suggests the presence of sequence-dependent asymmetric cooperativity in these regions \cite{orientationdependencereplication,directiondependencetranscription,repeattranscription}.

\begin{figure}
\begin{center}
\includegraphics[width=0.9\textwidth]{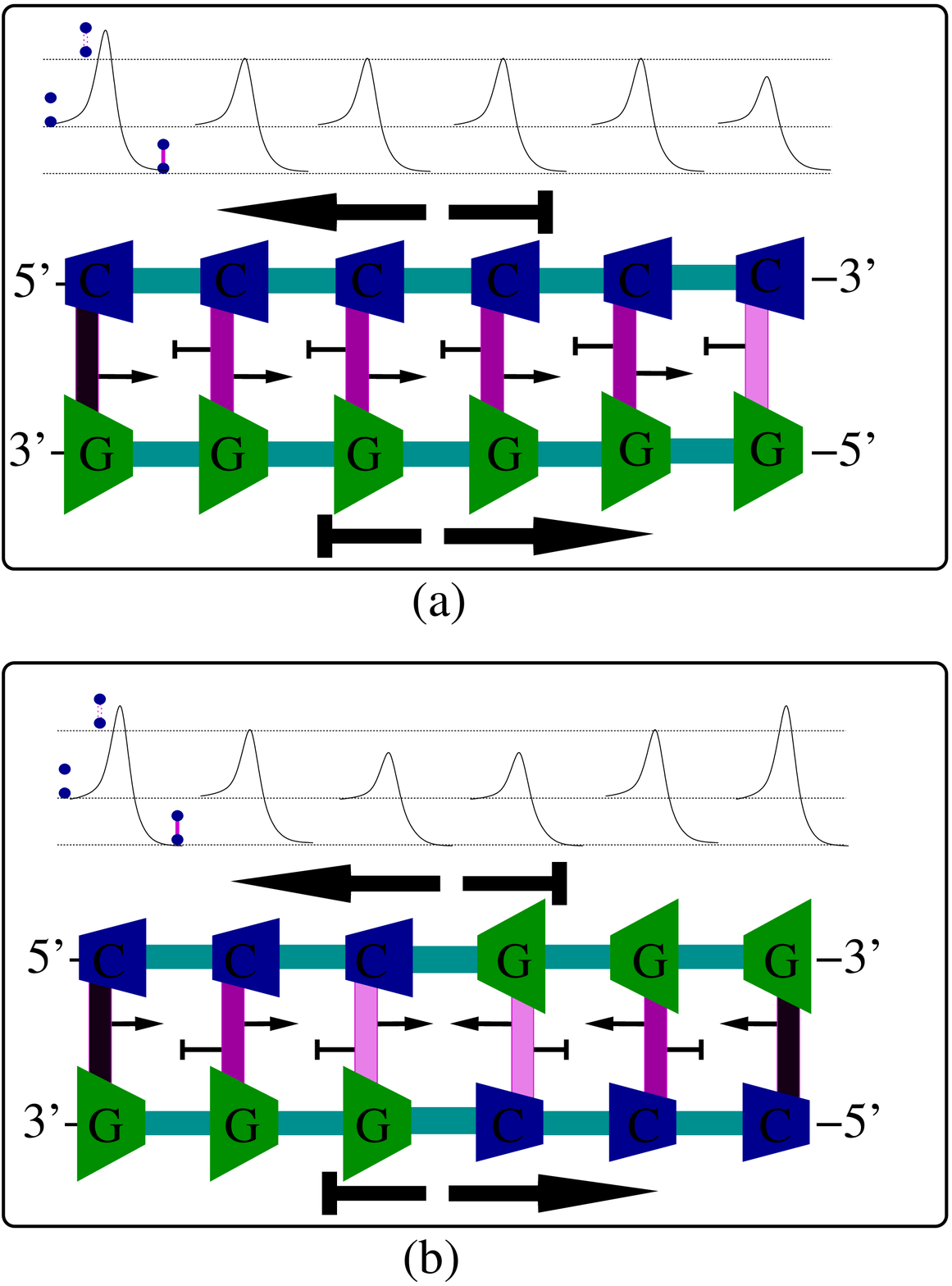}
\caption{Illustration of sequence-dependent asymmetric cooperativity (figure and caption adapted from \cite{scirep} under a Creative Commons Attribution $4.0$ International License). In a DNA double strand, the anti-parallel orientations of the two strands result in the cancellation of their respective opposing asymmetric cooperativity modes. If the nucleotides on both the strands are of the same type, the cancellation would be complete, due to symmetry. When the types of nucleotides on the two strands are different, say, with $C$ on the $3'-5'$ strand and $G$ on the $5'-3'$, the cancellation is not complete and the residual asymmetric cooperativity is dictated by the types of nucleotides, making the asymmetric cooperativity sequence-dependent. The thick arrows denote the sequence-independent asymmetric cooperativity dictated by the individual strands' directionality, whereas the thinner arrows attached to the hydrogen bonds denote the sequence-dependent asymmetric cooperativity that changes its mode depending on the orientation of the base-pair. $5' \hyp C \hyp 3'/3' \hyp G \hyp 5'$ base-pair orientation of the hydrogen bonds instantiates right asymmetric cooperativity, as shown in (a), whereas the $180^\circ$ -rotated $5' \hyp G \hyp 3'/3' \hyp C \hyp 5'$ instantiates left asymmetric cooperativity, as the last three bonds of (b) illustrates. Please note that the asymmetric cooperativity modes chosen here are opposite to that of our choice in \cite{scirep} (see text).The kinetic barrier diagrams above the strands in (a) and (b) are significantly different, illustrating the sequence-dependence of the unzipping behavior of DNA double strands. The unzipping of the strand in (a) would proceed sequentially from the rightmost end, whereas the middle two bonds would break and the strand will simultaneously unzip in both the directions in (b). This is proposed to lead to simultaneous construction of daughter strand on multiple segments of the single strand template in anti-parallel strands.}
\label{seqdependent}
\end{center}
\end{figure}

The observation of the extreme prevalence of GC skew or asymmetric nucleotide composition around replication origins in nearly all genomes studied thus far supports the presence of sequence-dependent asymmetric cooperativity in DNA \cite{rocha2004replication,replicationorientationskew,SPombeorigins,gcskeworigins1,gcskeweukaryotes,originsmammalsnatcomm}. In fact, most genomic analysis software programs use the switching of the sign of GC skew, defined as the running average of $(G-C)/(G+C)$ over large-enough windows of sequences, to identify the locations of replication origins \cite{oriloc}. The pervasive presence of GC (and/or AT) skews suggest that these skews serve an essential purpose of providing directionality for the DNA replication machinery, through asymmetric cooperativity, despite their presence being detrimental to the genome, since the skews reduce the amount of information that can be stored in the genome. There is also some evidence that the magnitude of GC skew is related to the speed of replication of genomes \cite{growthrategcskew}. In addition, it has been shown \cite{profoundskews, profoundskews2} that mammalian cells have profound GC and AT nucleotide skew switches near their replication origins, and that the magnitudes of skews correspond to the firing potential of the origins.  The significant difference between the observed mutation rates of $5' \hyp GC \hyp 3'$ (GpC) and $5' \hyp CG \hyp 3'$ (CpG) dinucleotides, where the latter undergoes substantially more single-strand-induced deamination and consequent transition from C to T, has been explained \cite{cpgmutationmelting, cpgtransitiongpc} as due to differences in the two dinucleotides' melting propensity. Our model provides a simple explanation for the difference in the melting propensity of the two dinucleotides, arising from the mutual stabilization of $5' \hyp GC \hyp 3'$ (GpC) dinucleotides due to asymmetric cooperativity, resulting in higher kinetic barriers and reduced melting, and the mutual weakening of kinetic barriers of the $5' \hyp CG \hyp 3'$ (CpG) dinucleotide. Our claim that the hydrogen bonds at the center of certain high-skew palindromic sequences have lower kinetic barriers due to sequence-dependent asymmetric cooperativity, as shown in fig. \ref{seqdependent}(b), resulting in them functioning as replication origins, is supported by the AFM-based experimental observation of local melting of replication origin of \textit{S.Pombe} genome \cite{originsmelting}. These low kinetic barriers render the palindromes susceptible to unwind locally, leading to cruciform extrusions \cite{palindromes}. This local instability is also connected to such palindromic sequences functioning as recombination hotspots \cite{palindromes}. Another line of evidence for our claim that high skews result in higher replicative potential comes from an \textit{in vitro} selection experiment \cite{selectionskews} where self-priming oligonucleotide replicators with different sequences were allowed to compete for limited monomer resources. The experiment showed that all the selected replicators showed maximum skews, with the template strand of the most fit sequences composed entirely of pyrimidines and the daughter strands, made of purines. However, because the employed sequences in this experiment were self-priming and initiated replication at the ends, these sequences were not palindromic. These experiments suggest that the unzipping behavior of DNA/RNA sequences are not dictated by thermodynamics alone, and asymmetric kinetic influences of the hydrogen bonds' neighborhood must be taken into consideration. More support for the presence of sequence-dependent asymmetric cooperativity in DNA is provided in our earlier paper \cite{scirep}.

Heteromolecular base-pairing, the base-pairing through hydrogen bonding of \textit{distinct} molecular species (purines with pyrimidines), is necessary for incorporation of sequence-dependent asymmetric cooperativity in duplex DNA. Homomolecular base-pairing cannot be used to incorporate asymmetric cooperativity due to the left-right symmetry of the homomolecular base-pairs. For example, the homomolecular base-pair $5' \hyp G \hyp 3'/3' \hyp G \hyp 5'$, and its $180^\circ$-rotated counterpart, $5' \hyp G \hyp 3'/3' \hyp G \hyp 5'$ are one and the same molecule, does not have left-right asymmetry, and hence cannot incorporate asymmetric cooperativity, which requires distinguishing between left and right directions \cite{scirep}. Hence a primordial heteropolymer with anti-parallel double strands supporting sequence-dependent asymmetric cooperativity can only be composed of \textit{even number of monomer species}.

From the foregoing, it should be clear that the location of the replication origins of DNA duplex and the rate and direction of its unzipping, which is the first step in the DNA replication process, is sequence-dependent, instantiated by GC and/or AT skews. This sequence-dependence of unzipping rates and origin locations, and hence of the replicative potential of DNA double strands (see below), makes certain DNA sequences evolutionarily dominant, among a pool of self-replicating DNA sequences, thereby selecting them. It should be noted that the asymmetric cooperativity model described above deals solely with the neighborhood dependence of \textit{kinetic barriers} for base-pair formation, and ignores other \textit{thermodynamic} aspects of DNA replication. The primary reason for this model choice is to highlight the \textit{sequence-dependence} of rate of replication, particularly the replicative competition between sequences with the same numbers of A, T, G and C, and hence a nearly constant thermodynamic unzipping potential (stacking interactions will contribute to some variability), but with distinct sequences that dictate their unzipping rates. Since the unzipping rates dictate the replicative potential of various sequences (see below), our assumption of the presence of asymmetric cooperativity allows us to introduce evolutionary competition between these sequences, and consequently, to the emergence of biological information through selection of specific sequences, even before the emergence of transcription and translation.

\subsubsection*{Model of replication - strand displacement model}

Self-replication of DNA/RNA (and their progenitors) involves both hydrogen-bond breaking to unzip the double strands and formation of hydrogen bonds between monomers and the template strand during the creation of the daughter strand. Since the reaction propensities of dissociation and formation of the hydrogen bonds are inversely related to each other, maximization of replication rate involves optimization with respect to parameters that affect both the hydrogen bond formation and dissociation rates, such as temperature, chemical potential, salinity, pH etc. Various replication schemes have been proposed in the literature to address the requirement of enabling both the formation and the dissociation of hydrogen bonds during self-replication of heteropolymers in the primordial Earth. 

One proposed scheme involves cyclical variation in the temperature around the melting temperature of the hydrogen bond, due to, say, day-night cycles, assisting in self-replication \cite{prernanatcomm}. At the high-temperature phase, the RNA/pre-RNA double strands would melt, resulting in single strands, which would then function as templates for daughter strand construction during the low-temperature phase. Chemical potential or salinity variations in periodically replenished tide-pools may have similarly helped self-replication. A major problem with this replication scheme is product inhibition \cite{rollingcirclereplication,nonenzymaticprimerextension} . Single-stranded templates reannealing with their complementary strands at low temperatures would inhibit daughter strand construction, leading to reduced replication potential. More over, the requirement of cyclical temperature/chemical potential variations in the environment for self-replication restricts the niches of the early life forms to regions with such variations, resulting in an evolutionarily suboptimal strategy.

Another scheme that addresses the aforementioned problem, and is universally implemented in the genomes of all non-viral life forms, is replication through strand displacement. This scheme obviates the requirement of variation of environmental parameters for self-replication. In extant genomes, replication happens isothermally under environmental conditions that favor hydrogen bond formation, which renders DNA double-strand unzipping (and hence the displacement of the complementary strand from the template) energetically unfavorable. Extant genomes employ ATP-dependent helicases to unzip the double-strands ahead of the DNA polymerase \cite{helicasereplication}, whereas, in the primordial scenario, thermal fluctuations and/or the energy released from polymerization of the daughter strand \cite{synthesizinglife,isothermalreplication} might have helped in unzipping the double strands of RNA/pre-RNA. Since this model of replication avoids the problems of product inhibition and the spatial constraints imposed by the requirement of cyclical environmental parameter variations, primordial self-replicators might have evolved to adopt the strand displacement scheme of self-replication, as is apparent from its universal usage across all non-viral life forms.

It is obvious that, within the strand displacement model, the \textit{rate-limiting step} for self-replication of DNA/RNA/pre-RNA is the double strand unzipping or the displacement of the complementary strand from the template strand, to make space for the nascent daughter strand construction \cite{isothermalamplification1,isothermalamplification2,isothermalamplification3}. In this paper, therefore, \textit{we equate the replicative potential of a primordial heteropolymer sequence to its unzipping rate}, in accordance with our assumption that the rate of replication dictated the replicative potential of primordial self-replicators.

In the following, we extend our premise of the presence of asymmetric cooperativity in DNA to RNA and pre-RNA (the evolutionary progenitor of RNA) as well, and will apply the above model of replication to show that the evolutionarily dominant early circular RNA/pre-RNA sequences are palindromes with maximal skews, and are thus predisposed to form hairpin or stem-loop secondary structures, one of the basic secondary structures of early catalytic RNA molecules \cite{hairpinintro}. We will then show that the evolutionarily dominant, highly skewed, \textit{binary-letter}, RNA/pre-RNA palindromic sequences can fold into many possible, nonspecific hairpin structures, whereas, the evolutionarily dominant, highly skewed, \textit{quadruplet-letter} palindromic sequences will fold into a single, specific hairpin secondary structure. In the following, we will simply use ``RNA'' to denote both RNA and/or any progenitors of RNA, for the sake of brevity, and our conclusions apply equally well for RNA's progenitors, as for RNA itself. This ambiguity in nomenclature is caused by our inability to precisely specify the self-replicator species in which the quadruplet letter alphabet arose first, and hence can only be referred to nonspecifically.

\section*{Results}

\subsubsection*{Palindromic sequences with high skews have higher replicative potential}
\begin{figure}
\begin{center}
\includegraphics[width=0.8\textwidth]{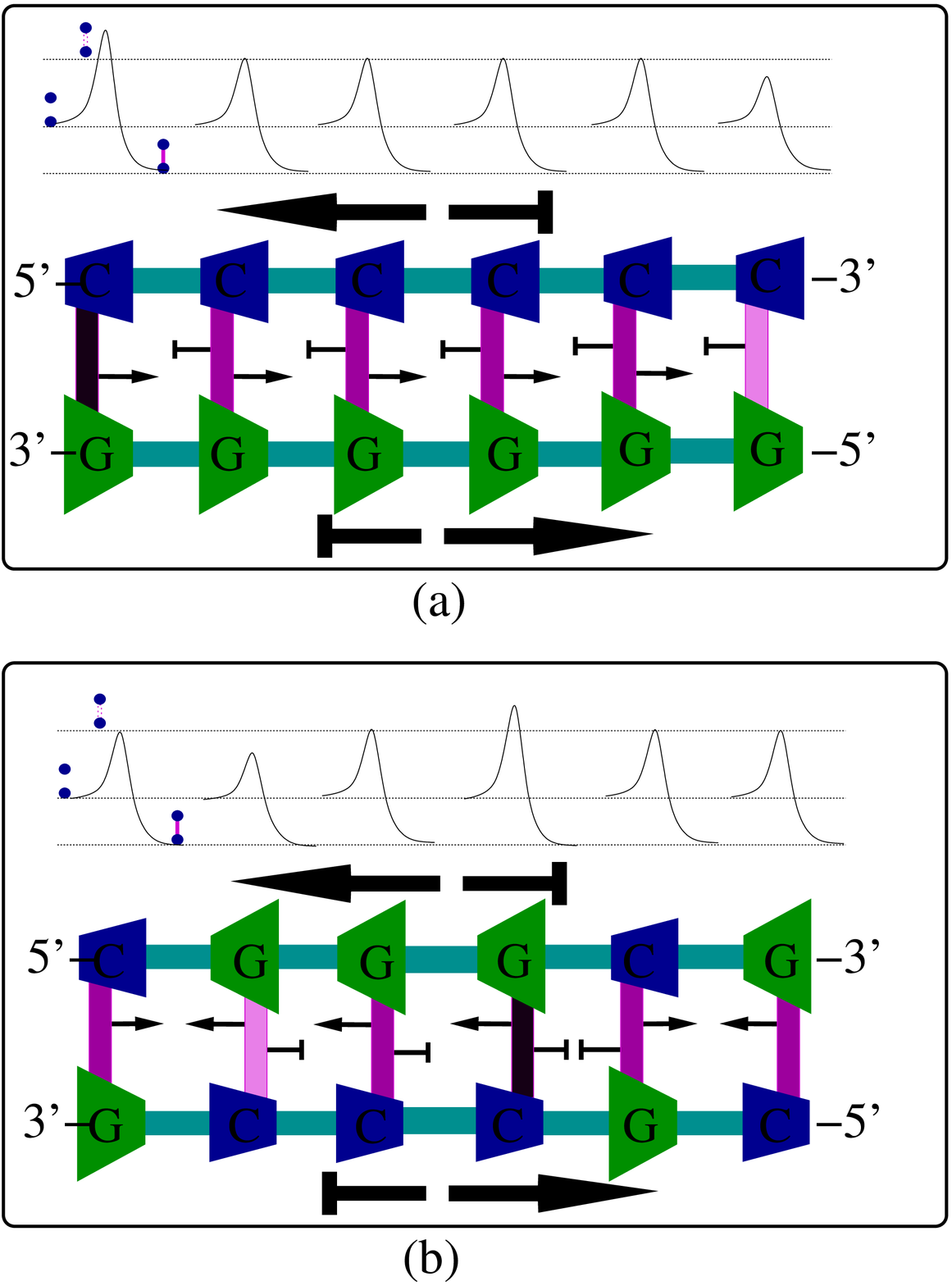}
\caption{Illustration of high kinetic barriers presented by binary-letter RNA with an arbitrary sequence. (a) A homogeneous sequence composed entirely of a single nucleotide in a binary-letter RNA double-strand $5' \hyp C_6 \hyp 3'/3' \hyp G_6 \hyp 5'$ is shown. This sequence presents smaller kinetic barrier for right-to-left unzipping and hence for the movement of the replisome through it from the right side. The kinetic barriers for individual hydrogen bonds are shown above the sequence. Although the kinetic barrier of the left-most bond is higher in the diagram, as the unzipping proceeds from the right, the barriers of bonds at the unzipping front become smaller. For example, the kinetic barrier of the second-to-last bond to the right is reduced when the last bond to the right is broken, due to the lack of stabilizing influence on the former by the latter. (b) An arbitrary sequence of a binary-letter RNA molecule presents a high kinetic barrier for the movement of the replisome through it, irrespective of the direction of the replisome movement. The high kinetic barrier arises from the mutually stabilizing influence of the $/5' \hyp GC \hyp 3'/3' \hyp CG \hyp 5'$ base-pairs, located at the fourth and fifth positions from the left of the sequence. This mutual stabilization makes breaking of the hydrogen bonds of the two base-pairs kinetically more unfavorable, irrespective of the direction of unzipping, compared to the barriers presented by the homogeneous sequence in (a). This renders such non-homogeneous binary-letter sequences replicatively inferior when compared with homogeneous sequences of same length. This demonstrates the necessity of homogeneous sequences for rapid unzipping in binary-letter RNA sequences.}
\label{gcmutualstable}
\end{center}
\end{figure}

Consider an RNA double-strand with a sequence of length $N$. We will assume that this RNA sequence is composed entirely of only two nucleotides, $G$ and $C$. To be concrete, we assume that the base-pair $5' \hyp G \hyp 3'/3' \hyp C \hyp 5'$ is left asymmetrically cooperative, which reduces the kinetic barrier of its left neighboring hydrogen bond and raises the barrier of its right neighbor. It is obvious that the base-pair $5' \hyp C \hyp 3'/3' \hyp G \hyp 5'$ would be right asymmetrically cooperative. The mode of sequence-dependent asymmetric cooperativity for RNA, i.e., whether $5' \hyp G \hyp 3'/3' \hyp C \hyp 5'$ catalyzes its left neighboring hydrogen bond or its right neighbor, has not been determined experimentally, which allows us the freedom to choose either of the modes to illustrate our results. We again emphasize that our choice of the asmmetric cooperativity mode does not affect our conclusions in any way below, and a specific choice is made solely for illustration purposes. An RNA single strand sequence composed of a random mixture of G's and C's cannot have maximal unzipping rate of all sequences, due to the presence of the mutually stabilizing dinucleotides $/5' \hyp GC \hyp 3'/3' \hyp CG \hyp 5'$, which presents a higher kinetic barrier for unzipping. This is demonstrated in fig. \ref{gcmutualstable}(b), where the fourth and the fifth nucleotides from the left mutually stabilize each other, thereby presenting a kinetic barrier. The linear sequence $5' \hyp C_N \hyp 3'/3' \hyp G_N \hyp 5'$ do not have the above dinucleotide that impedes rapid unzipping, and will unzip from its right-most end, thereby allowing replication to proceed sequentially from the right-most end to left-most end, as illustrated in fig. \ref{gcmutualstable}(a). The above homopolymeric sequence is still replicatively inferior to the palindromic sequence $5' \hyp C_{N/2}G_{N/2} \hyp 3'/3' \hyp G_{N/2}C_{N/2} \hyp 5'$, which initially unzips at its center, due to the reduced kinetic barrier at  the central $/5' \hyp CG \hyp 3'/3' \hyp GC \hyp 5'$ dinucleotide, and proceeds to unzip bidirectionally towards left and right, due to the permissive right and left asymmetric cooperativity modes of $5' \hyp C_{N/2} \hyp 3'/3' \hyp G_{N/2}\hyp 5'$ and $5' \hyp G_{N/2} \hyp 3'/3' \hyp C_{N/2} \hyp 5'$ replichores, respectively. We call the above sequence a maximally-skewed palindrome, since the nucleotide skew, defined as percentage of $(G-C)/(G+C)$ over running windows of fixed length, of the two strands of the duplex RNA on each replichore, is maximal in magnitude. Fig.\ref{seqdependent}(b) shows one such maximally-skewed palindrome. As can be seen in this figure, the two middle bonds have lower kinetic barriers, are weaker, and hence will be the first ones to break, thereby initiating the cooperative unzipping of the entire replichores on either side, similar to the observed earlier melting of replication origins \cite{originsmelting}. These middle bonds serve as an origin of replication. Thus the rate of unzipping of the entire palindromic sequence would be nearly twice that of the homopolymeric sequence above, thereby increasing its replicative potential. Moreover, if we confine our attention to circular genomes, the choice of genome topology of primitive organisms (prokaryotes), homopolymeric sequences such as $5' \hyp C_N \hyp 3'/3' \hyp G_N \hyp 5'$ are at a further disadvantage, because of the impossibility of formation of sequence-dependent origin of replication, since the asymmetric cooperativity mode would be the same for the entire length of the genome, with constant kinetic barriers. In the following, we restrict ourselves to this evolutionarily earlier and simpler circular topology for RNA self-replicator \cite{circularrna}, thereby avoiding the inherent problems associated with replicating the ends of linear replicators of recent evolutionary history \cite{telomerereplication}.

In order to improve the rate of unzipping of sequences of fixed length $N$, we can proceed as above to divide the self-replicator into shorter replichores, such as $5' \hyp C_{N/4}G_{N/4}C_{N/4}G_{N/4} \hyp 3'/3' \hyp G_{N/4}C_{N/4}G_{N/4}C_{N/4} \hyp 5'$, which would introduce two origins of replication and hence would be expected to replicate nearly twice as fast as the longer palindrome $5' \hyp C_{N/2}G_{N/2} \hyp 3'/3' \hyp G_{N/2}C_{N/2} \hyp 5'$. However, the former would consume monomers at a rate twice that of the latter, for self-replication, and hence, in environments where monomer supply is short, would not be favored. Sequence-dependent asymmetric cooperativity thus enables sequences to adapt to different environments that differ in their monomer supply rates. A relatively small, circular RNA self-replicator composed of a single palindrome and a single replication origin would have been replicatively more successful in the regime of limited monomer supply, when compared to both replicators with more than one replication origin and homogeneous-sequence replicators with no replication origin. The former are at, or soon would be at, a disadvantage due to low monomer supply, resulting in multiple replichores of the same replicator competing against each other for monomers. The latter, with no replication origin, cannot easily unzip from a predictable origin, due to high kinetic barriers across the length of the circular polymer. The foregoing is supported by the observation that the single-replication-origin genomes are the choice of most prokaryotes, that face the fiercest competition for monomeric resources, due to their numerosity.

Let us then concentrate on the single-replication-origin palindrome $5' \hyp C_{N}G_{N} \hyp 3'/3' \hyp G_{N}C_{N} \hyp 5'$, of length $2N$, of the order of tens of nucleotides, the size of an average stem-loop inverted-repeat sequence found in tRNA molecules \cite{trnastemlooplength}. Any modification of the sequence of this palindrome will adversely affect its smooth unzipping, due to switching of the sequence-dependent asymmetric cooperativity mode near the modified sections of the sequence. This mode-switching introduces high kinetic barriers, as illustrated in fig. \ref{gcmutualstable}, which in turn delays unzipping of sections of template for daughter strand construction, making such sequences replicatively inferior to the maximally-skewed palindromic sequences. The latter do not have such high kinetic barriers, and hence will be able to attract more monomers to form inter-strand hydrogen bonds with the exposed templates, and hence will be replicatively superior. From the foregoing, it becomes obvious that a \textit{circular RNA self-replicator composed of only two nucleotide letters can maximize its replicative potential only if its sequence is a maximally-skewed palidrome}, within our premise of the presence of sequence-dependent asymmetric cooperativity in RNA.

It has to be remembered that, in our model, we have restricted asymmetric cooperativity to operate only between nearest-neighbor inter-strand hydrogen bonds, for ease of analysis and illustration, whereas, in reality, it most probably extends over many inter-strand hydrogen bonds on either side \cite{nextnearestneighbor1,  longrangeeffectdna, longrangeeffectsdna2, electroniccoherence}. These longer-range interactions make the effect of high barriers resulting from any deviations from the maximally skewed palindromic sequence to be felt farther than nearest-neighbors, and hence can even reduce the unzipping potential of the origin of replication itself, rendering such sequences replicatively more inferior. Due to the longer range of asymmetrically cooperative interactions between base-pairs, the effective kinetic barrier reduction/enhancement at any single base-pair would be a weighted average of the effect of asymmetric cooperativity from an extended region, perhaps of the order of ten base-pairs. Such averaging can both dilute the deleterious effect of a single misoriented (left asymmetric cooperativity -moded in an otherwise right-moded sequence or vice versa) base-pair, and spread the deleterious effect over the entire extended region. A corollary is that, if a slim majority of base-pairs in an extended region have the same asymmetric cooperativity mode, the entire region will unzip cooperatively, albeit at a lower rate.


\subsubsection*{Binary letter maximally-skewed palindromes fold into nonspecific hairpin structures}
\begin{figure}
\begin{center}
\includegraphics[width=0.8\textwidth]{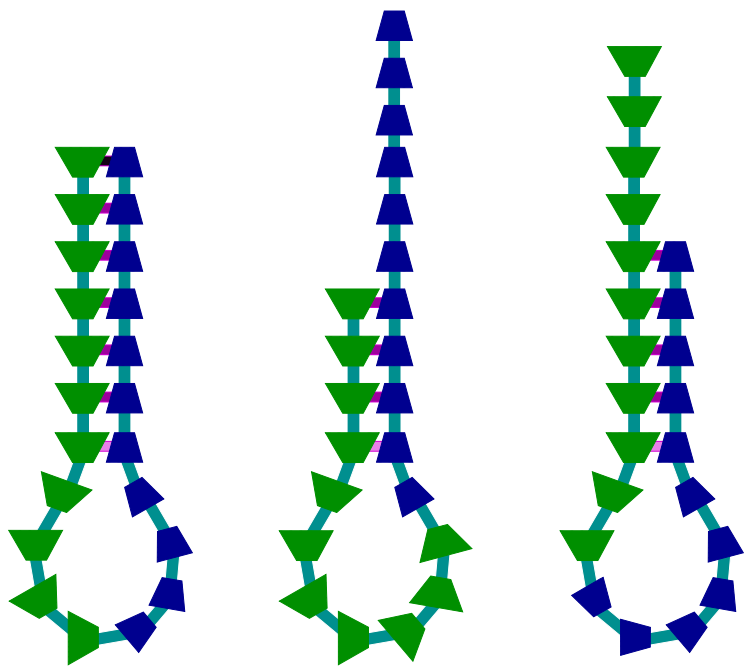}
\caption{This illustration shows a few possible secondary structures formed by a section of a circular, binary-letter, maximally skewed, palindromic RNA sequence. Due to the homogeneity of the sequences on either side of the origin of replication, multiple kinetically stabilized secondary structures are possible, leading to structural non-specificity or conformational plasticity. Bulges in the stem are excluded for simplicity. The capability of such non-specific secondary structures to catalyze the self-replication of their RNA templates is variable and sequence-independent, when compared to a quadruplet-letter, structurally specific, secondary structures, leaving the former at a selective disadvantage. Binary letter, maximally skewed, palindromic sequences cannot encode information about specific secondary structures that help catalyze their own or their hypercyclic partners' self-replication, due to the sequence homogeneity within each replichore. Quadruplet letter, maximally skewed sequences can simultaneously satisfy the unzipping requirements of replicative potential maximization and the secondary structure specificity for catalysis of self-replication.}
\label{manyconformations}
\end{center}
\end{figure}

Single-stranded maximally skewed palindromic sequences, such as $5' \hyp C_{N}G_{N} \hyp 3'$, can form stem-loop secondary structures. However, due to the homogeneous distribution of a single type of nucleotide within a single replichore of the above sequence, no information about specific secondary structures can be stored in the sequence, resulting in possibility of formation of multiple nonspecific secondary structures, leading to conformational plasticity. This is illustrated in fig.\ref{manyconformations}. The $5' \hyp C_{N}\hyp 3'$ replichore arm of the palindrome can bind with the $3' \hyp G_{N} \hyp 5'$ arm, to form a double strand in many possible ways,  due to the homogeneous nature of the sequence. For example, the palindrome can form an incomplete stem-loop structure, with a double-stranded stem segment of length $s$ amidst a loop segment of length $l$ and an overhang of length $h$. The lengths $s$, $l$ and $h$ are variable, within the constraint $2s+l+h = 2N$, quantifying the variety of seconary structures that can form from a binary-letter maximally-skewed palindromic single strand. We have excluded the possibility of bulges within the dsRNA for simplicity, above. Fig. \ref{manyconformations} shows three of the possible secondary structures. These nonspecific secondary structures will show variable catalytic activity due to the sequence variability in their loops. It has been known that secondary structures formed by RNA sequences with low information content, such as trinucleotide repeats, are nonspecific and form several alternative hairpin loop structures \cite{trinucleotiderepeatfolding}. A roughly similar argument has been made theoretically elsewhere, without invoking asymmetric cooperativity \cite{alphabetsizeprl,  alphabetsize2}. In \cite{ alphabetsize2}, it was theoretically shown that long, random RNA sequences can form secondary structures only when the number of letters are in the range between $2$ and $4$.

It can be claimed that among all nonspecific secondary structures the above palindromic sequence is allowed to adopt, only a few that minimize the free energy substantially will actually form, making the above argument against binary letter genomes invalid. However, it has been demonstrated that rather than large free energy gap between ground state and other suboptimal configurations, the number of nucleation centers and the exact folding pathways dictate the folding behavior of RNA secondary structures, specifically in simple RNA hairpins \cite{rnafoldingkinetics,rnahairpin,rnaenergylandscape}. The ``thermodynamic hypothesis'' that RNA single strands always fold into a structure that globally minimizes the free energy has been shown to be grossly inaccurate, and the folding landscape is now thought to be rugged, with multiple possible configurations allowed, corresponding to local minima in the landscape that are separated by large kinetic barriers \cite{multiplenativestates}. Despite their small size, even tRNAs fold into distinct configurations with different biological activities \cite{trnaconformations,rnasecondarystructure,fitnesslandscape}. Therefore, minimization of free energy to obtain stable secondary structures cannot instruct us on other possible functionally relevant secondary structures that are kinetically stabilized \cite{kineticstability}, making such an analysis irrelevant for our current purposes.

The foregoing illustrates the conflict between high replicative potential and information-storing capability in a binary-letter self-replicator. Whereas the replicative potential is maximized by a single low-entropy sequence in a binary-letter RNA, the ability to store information requires a relatively large sequence space with a relatively flat fitness landscape, thereby allowing the evolutionary selection process to operate. Introduction of another pair of nucleotides in the RNA self-replicator system simultaneously satisfies both the above constraints.  



\subsubsection*{Secondary structure specificity requires quadruplet letter sequences}
We have laid out above the conflict between information-storing capability and the replicative potential of binary-letter RNA sequences, within the premise of the presence of sequence-dependent asymmetric cooperativity. Since the replicative advantage provided by high-skew palindromic sequences cannot be foregone without jeopardizing evolutionary superiority over other sequences, evolution must have found a way around the limit imposed by the near-homogeneous, evolutionarily superior, binary-letter sequences on information storage, by selecting alphabets with larger number of letters. The number of letters must increase by two, in order to satisfy the constraint of left-right asymmetry, required to incorporate sequence-dependent asymmetric cooperativity, as explained above, and also in \cite{scirep}; i.e., the base-pair formation must be between two distinct monomers, in order to incorporate sequence-dependent asymmetric cooperativity. Thus the minimal number of letters needed to simultaneously satisfy the requirements of superior replicative potential and information-storing potential becomes \textit{four}. With four letters, all sequences composed only of, for e.g., A and G nucleotides (purines) on one single-strand arm of the palindrome, i.e., on one replichore, and U and C (pyrimidines) on the other, would retain their high-skew character and asymmetric cooperativity mode, and hence would still have high replicative potential, assuming similar sequence-dependent asymmetric cooperativities for both the GC and AT base-pairs. It has to be noted that, with the introduction of two more letters, the constraint that maximally skewed sequences should be palindromic can also be relaxed, for instantiation of replication origins. This increase in the number of letters significantly enlarges the subset of the maximum-skew sequences whose replicative potential is maximum, and allows evolution to select sequences with superior catalytic ability within this subset. 

For example, the five-nucleotide double-stranded sequences $5' \hyp UCUUC \hyp 3'/3' \hyp AGAAG \hyp 5'$ and $5' \hyp  CUCUU 3'/3' \hyp GAGAA \hyp 5'$ will have similar unzipping kinetics due to the same mode of sequence-dependent asymmetric cooperativity in both, with all base-pairs catalyzing their right-neighboring hydrogen bonds. On the other hand, the sequence $5' \hyp UGCUU \hyp 3'/3' \hyp ACGAA \hyp 5'$ will have significantly lower unzipping rate compared to the former due to the mutually stabilizing influence of the $5' \hyp GC \hyp 3'$ base-pairs \cite{cpgmutationmelting, cpgtransitiongpc}, arising from the switching of the local sequence-dependent asymmetric cooperativity mode, which also reduces the GC skew magnitude of the sequence (see fig.\ref{gcmutualstable}(b)). Whereas $5' \hyp GGGGG \hyp 3'$ is the only possible sequence that is maximally skewed within the subset of all binary-letter five-nucleotide long RNA sequences, a quadruplet letter five-nucleotide long RNA can have $32$ possible maximally-skewed sequences, where each of the five positions can take either a G or an A (a purine), while preserving the sequence-dependent asymmetric cooperativity mode and hence the unzipping kinetics, and consequently, the replicative potential. This enlarged subset of sequences, with a nearly-flat replicative potential, allows evolution to explore sequences with catalytic abilities that enhances the rate of self-replication of themselves and/or its hypercyclic partners, without paying any fitness penalty. Inclusion of the difference in the dissociation rates between AU and GC hydrogen bonds will reduce the size of this subset and alter the replicative potential landscape, but it would still be larger than the binary-letter subset of size one.

\begin{figure}
\begin{center}
\includegraphics[width=0.25\textwidth, angle=90, origin=c]{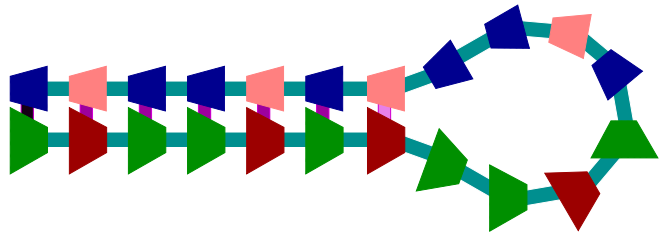}
\caption{Illustration of the specificity of the secondary structure formed by a section of a circular, quadruplet-letter, maximally skewed, palindromic RNA sequence. With four letters, the sequence of a section of the RNA self-replicator completely specifies its secondary structure. Evolution can modify the sequence for an appropriate secondary structure to enhance its catalytic potential, without jeopardising the sequence's replicative potential, assuming that both the base-pairs have similar asymmetric cooperativities. Such an evolutionary modification is not possible in a binary-letter, maximally-skewed sequence. Placing purines on one and the pyrimidines on the other replichore of a quadruplet letter sequence, with the two replichores separated by the origin of replication, satisfies the maximal-skew constraint required for easy bi-directional unzipping from the origin, and also provides two nucleotides on each replichore to constrain the possible secondary structures that the sequence can adopt. This evolutionary advantage provided the quadruplet-letter sequences superiority over binary-letter sequences, and possibly offset the disadvantages of the complex production mechanisms of all the four monomers required by the former, within our model.}
\label{singleconformation}
\end{center}
\end{figure}

With four letters, the replicatively successful maximally skewed RNA sequences can also fold into catalytically active, \textit{specific} secondary structures, such as stem-loops. This is illustrated in the schematic fig. \ref{singleconformation}.  Whereas the secondary structures formed by single-stranded maximally-skewed \textit{binary-alphabet} sequences, such as $5' \hyp C_{N}G_{N} \hyp 3'$, are constrained to be nonspecific due to sequence homogeneity, the secondary structures that can be formed by maximally-skewed \textit{quadruplet letter} sequences are highly sequence-specific, due to the specificity of base-pairing between G and C, and between A and U (and, to a certain extent, between G and U). Unlike in the binary letter case, the lengths of the loop $l$, double-stranded region $s$ and of the single-stranded overhang $h$ of a stem-loop secondary structure can be specified completely by the quadruplet letter RNA sequence, allowing evolution to select the sequences that produce catalytically active secondary structures that are useful for self-replication. This selection may have operated either on a single such sequence, or more probably, on a set of sequences that are hypercyclically coupled together, forming a quasispecies \cite{eigen}.

\subsection*{Why aren't the extant genomes maximally skewed in nucleotide composition?}

We have argued above that the evolutionarily successful sequences of the RNA world that were simultaneously self-replicating and autocatalytic must have been maximally-skewed quadruplet letter sequences that were capable of forming structurally-specific hairpin loop secondary structures, within our premise that asymmetric cooperativity is present in RNA. Beyond the replication origins, where the skews are substantial \cite{profoundskews, profoundskews2}, we don't generally find such maximally-skewed sequences elsewhere in extant RNA/DNA genomes. What mitigates the need for high skew in these regions in extant genomes?

In case of free-floating RNA strands, replicative potential is the only input to calculate the evolutionary fitness function, and maximal-skew sequences are beneficial, as argued above. However, the fitness function of a more evolved organism depend on traits beyond replication rate, and includes information in its genome about extracting energetic and material resources from its environment, which manifests through transcription. Since high GC or AT skew will reduce the number of nucleotides of a specific type (say, pyrimidines) on the coding strand, the codons with those nucleotides cannot be used to encode for amino acids, resulting in reduced ability to store information in the genome. Thus, an increased pressure to store information in extant organisms may have exerted a downward pressure on the magnitude of nucleotide skews in extant genomes. Indirect evidence for this claim comes from the observation that the skews in the coding sequences of various genomes is higher in the third codon position, when compared to the first two positions, where the evolutionary pressure from information storage is higher \cite{lobry}. Transcribed ribozyme sequences that need to form stem-loop secondary structures to perform their catalytic functions, cannot support high skews, due to the base-pairing constraint between the two arms of the stem sequence, which requires equal numbers of purines and pyrimidines in the sections coding for stems. The evolutionary pressure arising from this constraint also reduces the magnitude of the overall skew in tRNA- and rRNA-coding sequences, and might have operated even in the RNA-world scenario. This downward pressure on the nucleotide skews arising from the need for information storage, together with an upward pressure to provide directional signals for transcription and replication machineries, possibly sets the effective skews in extant genomes.


\subsubsection*{Falsification approaches}

Our arguments for the need for quadruplet letter alphabet above are based on two theoretical predictions: (a) Maximally-skewed sequences, with pyrimidines occupying one half of the sequence (say, left) and their corresponding base-pairing purines occupying the other half (say, right, depending on the mode), will have the highest replication rate of all sequences of the same length, in the limit of low monomer concentration. (b) Replicatively superior hairpin or stem-loop secondary structures formed by quadruplet letter sequences have far fewer kinetically-stabilized, long-lifetime, secondary structure configurations, compared to binary letter sequences. Both these predictions are eminently testable. The prediction (a) can be tested in an \textit{in vitro} evolutionary selection experiment, as has been done here \cite{selectionskews} but without the self-priming from the sequence ends, to find sequences that are replicatively superior. Prediction (b) can be tested by carefully evaluating the impact of multiple conformers of a single high-skew, palindromic RNA sequence, composed of either binary or quadruplet letter alphabets, on NMR spectra, fluorescence spectroscopy, gel electrophoresis, or other novel secondary structure isolation/determination techniques.

\section*{Summary and Conclusion}
We have shown above that the replicatively superior, binary-letter self-replicators in the RNA-world were maximally-skewed, circular, palindromic sequences, capable of forming stem-loop secondary structures, within our premise of the presence of sequence-dependent asymmetric cooperativity in RNA. This superiority arises from the reduced kinetic barrier in the central region of the sequence, due to opposing modes of asymmetric cooperativity from either side, thereby creating an origin of replication there. This origin allows for simultaneous replication of the two arms of the palindromic sequence, resulting in reduced replication time. On the other hand, the constraint of maximal nucleotide skew with opposite signs in the two arms of the palindromic sequence, required for replicative potential maximization, severely reduces the information-storing capacity of the binary-letter inverted repeat sequences. This maximal skew constraint forces each arm of the binary letter inverted repeat to use just one nucleotide, thereby nullifying its information storage potential. We showed that the conflicting sequence requirements for both the maximization of replicative potential and information-storing capacity, the latter measured in terms of specificity of secondary structures that such RNA sequences can form, can be simultaneously satisfied only by sequences composed of a minimum of four-letter alphabet. This might explain the choice of quadruplet letter alphabet in the genomes of extant organisms.


\subsection*{Author Contributions}
HS conceived, researched and wrote the article. 

\subsection*{Declaration of Interests}
The author declares no competing interests.




\subsection*{Funding}
Support for this work was provided by the Science \& Engineering Research Board (SERB), Department of Science and Technology (DST), India, through a Core Research Grant with file no. CRG/2020/003555. 

\subsection*{Acknowledgements}  
I thank Robert A. Gatenby for providing the initial impetus for the above research and for his unwavering support.

\clearpage

\bibliographystyle{ieeetr}
\bibliography{quadruplet_citations}


\end{document}